\documentstyle[aaspp4,psfig]{article}

\begin{document}

\title{Do Globular Clusters Harbor Black Holes?}

\author{X.-Z. Zheng}
\affil{National Astronomical Observatories, Chinese Academy of Sciences, Beijing 100012\\ Email: zxz@alpha.bao.ac.cn}

\begin{abstract}
It has been firmly established that there exists a tight correlation 
between the central black hole mass and velocity dispersion 
(or luminosity) of elliptical galaxies, ``pseudobulges'' and 
bulges of galaxies, although the nature of this correlation still 
remains unclear. In this letter, we explore the possibility of
extrapolating such a correlation to less massive, spherical systems 
like globular clusters. In particular, motivated by the apparent 
success in globular cluster M15, we present an estimate of the 
central black hole mass for a number of globular clusters with 
available velocity dispersion in the literature.
\end{abstract}


\section{Introduction}

  Supermassive black holes have been convincingly detected in the centers 
of some nearby galaxies (Kormendy \& Richstone 1995). Kormendy \& Gebhardt 
(2001, hereafter KG2001) gave a comprehensive review 
of recent black hole discoveries made with the {\it Hubble Space Telescope} 
(HST). A tight correlation between black hole mass and bulge velocity 
dispersion is confirmed. They noticed that black hole mass correlates with 
the luminosity of ``pseudobulges'' in disk galaxies, elliptical 
galaxies and the bulges of disk galaxies, but is independent of the 
luminosity of galaxy disks. 
The correlation  strongly suggests a causal connection between the 
formation and evolution of the black hole and the bulge but the nature 
of this connection remains unknown.  

  It is interesting to check whether this correlation still applies to both
larger and smaller spherical systems.  We may even speculate whether the 
correlation extends to systems with dispersion as low as that of globular 
clusters, since the globular clusters are also self-gravitating spherical 
systems similar to galactic bulges. Using the M$_{BH}$ -- $\sigma$ 
correlation, a crude estimate for the possible black hole mass in globular 
clusters can be obtained. For a typical massive globular cluster having a 
dispersion of the order of 10 km\,s$^{-1}$, a black hole mass of about 
2$\times\,10^3$\,M$_\odot$ is expected.  In certain galaxies, the black 
hole directly reveals itself through its associated accretion and activity. 
Such activity can hardly happen in globular clusters due to 
shortage of gas. However, recent X-ray observations of several starburst
galaxies (e.g. M82, NGC 4038/39) reveal the existence of intermediate-mass
black hole in starburst regions which are related to the formation of 
globular clusters (Kaaret et al. 2001; Matsumoto et al. 2001; Fabbiano, Zezas,
\& Murray 2001).  The presence of a black hole in a globular cluster 
affects the stellar density profile and the central stellar dynamics. 
With the dynamical detection sensitivity currently available, black hole mass
as low as 1000 M$_\odot$ can hardly be identified (van der Marel 2001).
So far, the only example was presented by Gebhardt et al. (2000) for M15
which may possibly host a black hole of the order of 10$^3$ M$_\odot$.

\section{Constraint on M$_{BH}$ -- $\sigma$ correlation by M15}

Gebhardt et al. (2000) inferred the projected velocity dispersion profile from
$\sim$1800 member stars in M15 with known line-of-light velocities. Assuming
isotropic velocity distribution, a constant stellar  mass-to-light ratio 
(M/L)$_V$\,=\,1.7 and no rotation, the best spherical dynamical model 
that matches the data contains a 2000\,M$_\odot$ black hole (see fig.15 in 
Gebhardt et al. 2000). However, other models also explain the data, such as
the models presented by Gebhardt et al. (2000).
To draw a firm conclusion on this issue, further studies are needed.

  If the globular clusters are eventually found to possess central black 
holes, as hinted in Gebhardt et al. (2000), it is interesting to 
investigate the correlation between black hole mass and velocity dispersion 
among globular clusters. Taking the projected velocity dispersion at 
effective radius 9.0\,$\pm$\,0.7\,km\,s$^{-1}$ and assuming a black hole 
with mass of 2000\,M$_\odot$, M15 can be plotted into the M$_{BH}$ -- 
$\sigma_e$ diagram. In Fig. 1, the solid line defined by
\begin{displaymath}
   M_{BH}=(1.26\pm0.07)\,\times\,10^8\,M_\odot\,\left( \sigma_e \over {200\,{\rm km\,s}^{-1}} \right)^{3.42\pm0.08},
\end{displaymath}
is the best fit to the galaxies.  
Within the uncertainties, M15 is perfectly sitting on the lower
extension of the fit! Adding M15 to KG2001 data, a robust  M$_{BH}$ -- 
$\sigma_e$ correlation in a much larger range can be given as
\begin{displaymath}
   M_{BH}=(1.27\pm0.07)\,\times\,10^8\,M_\odot\,\left( \sigma_e \over {200\,{\rm km\,s}^{-1}} \right)^{3.50\pm0.04},
\end{displaymath}
which is shown by dash line in  Fig. 1. 
Should the globular clusters and galaxies follow similar physical process of black hole formation, they would exhibit the same correlation between black hole mass and velocity dispersion. Based on the data of globular clusters and galaxies, the determination of the M$_{BH}$ -- $\sigma$ correlation would be improved significantly.

\begin{figure}[ht]
\plotone{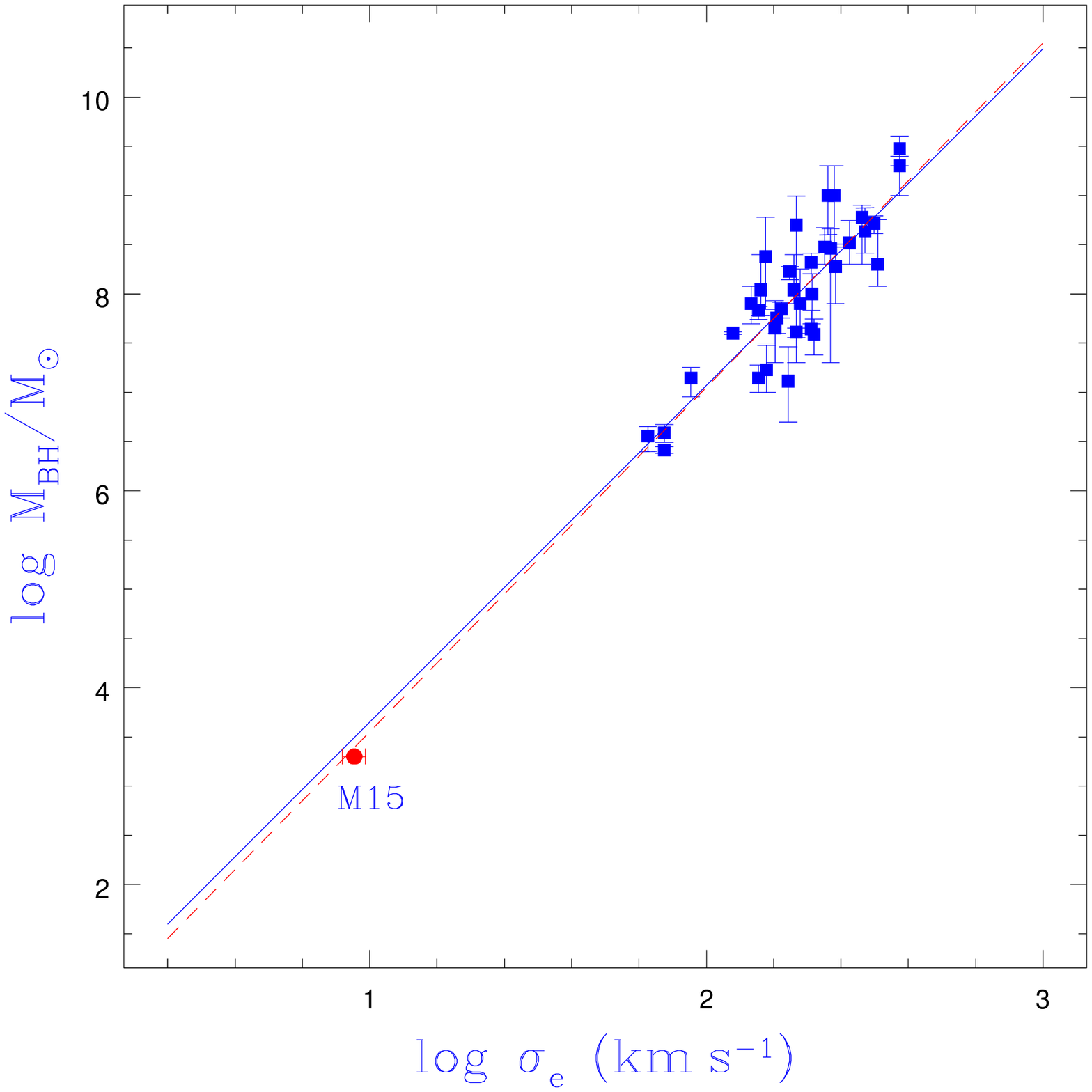}
\caption{Correlation of black hole mass with the mean velocity dispersion 
within the effective radius of the bulge of galaxies (squares, KG2001).
The globular cluster M15 (solid circle) possibly contains a black hole of 
2000\,M$_\odot$, consistent with the correlation derived from galaxies. 
}
\label{figm15}
\end{figure}

\begin{figure}
\plotone{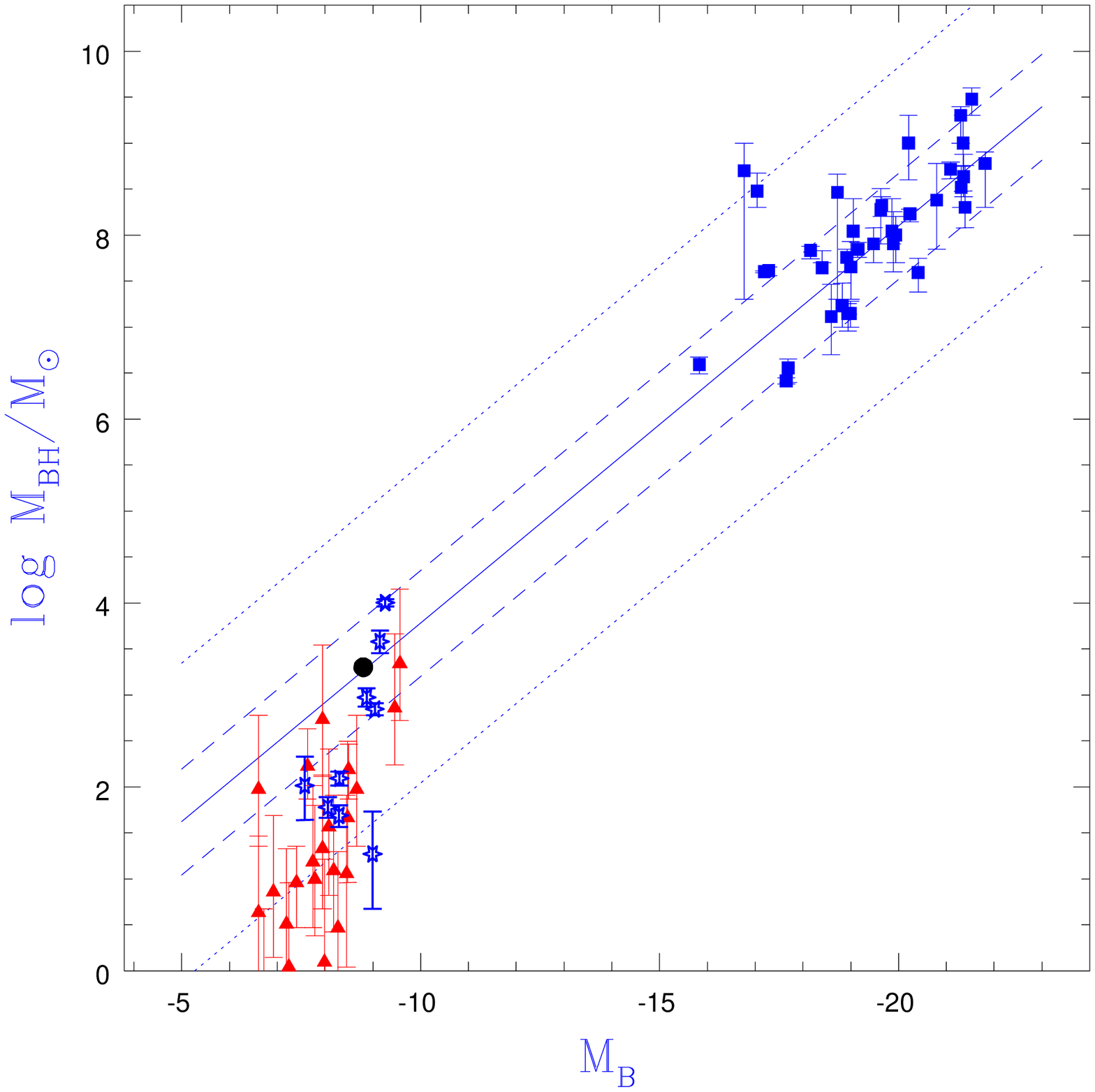}
\caption{Correlation of black hole mass with the absolute B-band magnitude 
for the bulge component of the host galaxy (squares, KG2001) and globular 
clusters (triangles for Galactic ones and asterisks for ones in M31).
The circle is M15. The solid line gives the best fit of galaxies
by KG2001. The dash lines and dot lines show 1\,$\sigma$ and 3\,$\sigma$ 
confidence limits respectively.
}
\label{figgc}
\end{figure}

\section{Black Hole Mass Estimate in Globular Clusters}

  Bearing in mind that black hole mass correlates only with velocity 
dispersion or luminosity of the self-gravitating spherical systems, a bold
speculation arises that the formation and growth of black hole can be linked
with a certain potential physical process that is universal for such systems,
and hence various sizes of self-gravitating systems satisfy the same 
correlation.  With this speculation, the same M$_{BH}$ -- $\sigma$ 
correlation should be found in a wide range of dimensions going from 
stellar globular clusters to galaxies. The black hole mass measurement
by Gebhardt et al. (2000) shows that the globular cluster M15 is 
remarkably consistent with the M$_{BH}$ -- $\sigma$ correlation derived 
from galaxies.  This implies that the speculation might be applicable at least
for globular clusters. 

  Adopting such a universal M$_{BH}$ -- $\sigma$ correlation, the mass of
the black hole in globular clusters can be estimated in terms of
velocity dispersion. In Table 1, the parameters of 22 Galactic 
and 9 M31 globular clusters are tabulated.
According to the relation between the black hole mass and central velocity 
dispersion of galaxies, given by equation (5) in Merritt \& Ferrarese (2001),
the mass of black holes in those globular clusters is estimated. The results 
areand listed in column (6) of Table 1. The uncertainties are given following 
those of velocity dispersion data. 

  For galaxies, black hole mass also correlates with bugle luminosity.
Such a correlation is expected yet to present in globular clusters. 
The luminosity of globular clusters  is plotted against black
hole mass in Fig. 2.  The solid line gives the best fit of galaxies
by KG2001. The globular clusters fall around the solid line
with a considerable scatter. This consistency reinforces the possibility that
the globular clusters harbor black holes and satisfy the same M$_{BH}$ -- 
$\sigma$ correlation as that of galaxies.

\begin{table}[t]
\caption[]{parameters of some globular clusters}
\label{table1}
\begin{center}
\begin{tabular}{cccrcccr}
\hline\hline\noalign{\smallskip}
   ID name & M$_{B}$ & $\sigma_0$ & logM$_{BH}$ & ID name &  M$_{B}$ & $\sigma_0$ & logM$_{BH}$ \\
 (1) & (2) & (3) & (4) &  (1) & (2) & (3) & (4) \\
\noalign{\smallskip}
\hline
\noalign{\smallskip}
\multicolumn{8}{c}{Galactic Globular Clusters\tablenotemark{\rm a}}\\
 NGC 104  & $-$8.66 &  10.0$^{+4.8}_{-2.6}$&  1.97$^{+0.80}_{-0.62}$ & NGC 6266 & $-$9.46 &  15.4$^{+7.4}_{-4.0}$&  2.86$^{+0.80}_{-0.62}$  \\
 NGC 362  & $-$7.78 &  6.2$^{+3.0}_{-1.6}$ &  0.99$^{+0.81}_{-0.61}$ & NGC 6284 & $-$7.75 &  6.8$^{+3.4}_{-2.0}$ &  1.18$^{+0.83}_{-0.71}$  \\
 NGC 1851 & $-$7.63 &  11.3$^{+2.5}_{-1.8}$&  2.22$^{+0.41}_{-0.36}$ & NGC 6293 & $-$8.08 &  8.2$^{+4.2}_{-2.5}$ &  1.57$^{+0.85}_{-0.75}$  \\
 NGC 1904 & $-$7.24 &  3.9$^{+2.2}_{-1.9}$ &  0.04$^{+0.92}_{-1.37}$ & NGC 6325 & $-$8.45 &  6.4$^{+3.3}_{-2.5}$ &  1.06$^{+0.85}_{-1.02}$  \\
 NGC 5272 & $-$8.27 &  4.8$^{+2.4}_{-1.4}$ &  0.47$^{+0.83}_{-0.71}$ & NGC 6342 & $-$6.61 &  5.2$^{+2.6}_{-1.5}$ &  0.63$^{+0.83}_{-0.70}$  \\
 NGC 5286 & $-$8.47 &  8.6$^{+4.3}_{-2.5}$ &  1.66$^{+0.83}_{-0.70}$ & NGC 6441 & $-$9.56 &  19.5$^{+9.4}_{-5.1}$&  3.34$^{+0.81}_{-0.62}$  \\
 NGC 5694 & $-$7.40 &  6.1$^{+1.3}_{-1.3}$ &  0.96$^{+0.40}_{-0.49}$ & NGC 6522 & $-$7.95 &  7.3$^{+3.5}_{-2.0}$ &  1.33$^{+0.80}_{-0.66}$  \\
 NGC 5824 & $-$8.49 &  11.1$^{+1.6}_{-1.6}$&  2.19$^{+0.28}_{-0.32}$ & NGC 6558 & $-$6.71 &  3.5$^{+1.8}_{-1.2}$ & $-$0.18$^{+0.85}_{-0.86}$  \\
 NGC 5904 & $-$8.18 &  6.5$^{+3.2}_{-1.8}$ &  1.09$^{+0.82}_{-0.66}$ & NGC 6681 & $-$6.61 &  10.0$^{+4.8}_{-2.6}$&  1.97$^{+0.80}_{-0.62}$  \\
 NGC 5946 & $-$7.98 &  4.0$^{+2.9}_{-2.9}$ &  0.09$^{+1.12}_{-2.65}$ & NGC 6752 & $-$7.19 &  4.9$^{+2.4}_{-1.4}$ &  0.51$^{+0.82}_{-0.69}$  \\
 NGC 6093 & $-$7.95 &  14.5$^{+7.0}_{-3.8}$&  2.73$^{+0.81}_{-0.62}$ & NGC 7099 & $-$6.92 &  5.8$^{+2.9}_{-1.7}$ &  0.86$^{+0.83}_{-0.71}$  \\

\noalign{\smallskip}                                                           
\hline
\noalign{\smallskip}                                                            
\multicolumn{8}{c}{ Globular Clusters in M31\tablenotemark{\rm b}} \\

 G58  & $-$8.30 & 10.6$\pm$ 0.4 & 3.46$^{+0.06}_{-0.01}$ &G219 & $-$9.00 & 7.1 $\pm$ 1.8 & 2.82$^{+0.36}_{-0.08}$  \\
 G73  & $-$9.04 & 15.3$\pm$ 0.5 & 4.04$^{+0.05}_{-0.01}$ &G272 & $-$8.87 & 16.3$\pm$ 0.8 & 4.14$^{+0.08}_{-0.01}$  \\
 G105 & $-$7.58 & 10.2$\pm$ 1.7 & 3.40$^{+0.24}_{-0.04}$ &G280 & $-$9.26 & 26.9$\pm$ 0.5 & 4.93$^{+0.03}_{-0.00}$  \\
 G108 & $-$8.29 & 8.7 $\pm$ 0.5 & 3.14$^{+0.09}_{-0.02}$ &G319 & $-$8.06 & 9.1 $\pm$ 0.5 & 3.22$^{+0.08}_{-0.01}$  \\
 G213 & $-$9.15 & 21.9$\pm$ 1.3 & 4.61$^{+0.09}_{-0.01}$ & & & & \\

\noalign{\smallskip}
\hline
\hline
\noalign{\smallskip}
\end{tabular}
\end{center}
\begin{footnotesize}
$^{\rm a}$column (2): Harris 1996; column (3): Dubath et al. (1997).\\
$^{\rm b}$column (2) -- column (3): Dubath \& Grillmair (1997).\\
(1) Name of globular cluster;
(2) Absolute B-band magnitude (cluster luminosity);
(3) Projected central velocity dispersion in km\,s$^{-1}$;
(4) Logarithm of black hole mass in solar mass.
 
\end{footnotesize}
\end{table}
  
\acknowledgments
   It is a pleasure to thank Prof. X.-P. Wu for his creative advice and 
discussions. The author is extremely grateful to Prof. L.-C. Deng for 
excellent suggestions and so much help. Many thanks to the staffs of stars 
and stellar systems group for their very kind help. The author also thank 
the referee Prof. Z.-G. Deng for helpful suggestions that improve the paper.
This work is supported in part by the Ministry of Science and Technology of 
China through grant G19990754.

\end{document}